\documentclass[prd,aps,nofootinbib,twocolumn,showpacs,amsmath,superscriptaddress]{revtex4-2}
\usepackage{graphicx}
\usepackage{comment}
\usepackage{epstopdf}
\usepackage{bm}
\usepackage{epsfig}
\usepackage{graphics}
\usepackage{xspace}
\usepackage{amssymb}
\usepackage{amsmath}
\usepackage{xcolor}
\usepackage[utf8]{inputenc}
\usepackage[normalem]{ulem}
\usepackage{stmaryrd}
\usepackage[colorlinks=true,allcolors=blue]{hyperref}
\usepackage{caption}
\usepackage{subcaption}
\def\OMIT#1{}

\newcommand{\nn}{\nonumber}

\newcommand{\bea}{\begin{eqnarray}}
\newcommand{\eea}{\end{eqnarray}}

\newcommand{\beq}{\begin{equation}}
\newcommand{\eeq}{\end{equation}}

\newcommand{\gsim}{\mathrel{\rlap{\lower4pt\hbox{\hskip1pt$\sim$}}\raise1pt\hbox{$>$}}}

\newcommand{\be}{\begin{equation}}
\newcommand{\ee}{\end{equation}}

\begin{document}

\title{\bf Universality and Kinematic Dependence of Hadronization Effects  \\ in DIS Global Event Shapes}

\author{Radja Boughezal}
\email{radja.boughezal@gmail.com}
\affiliation{Northwestern University, Evanston, IL, USA}

\author{Haotian Cao}
\email{haotiao.cao@northwestern.edu}
\affiliation{Northwestern University, Evanston, IL, USA}
\affiliation{Center for Frontiers in Nuclear Science, Stony Brook University, Stony Brook, NY 11794, USA}

\author{Zhong-Bo Kang}
\email{zkang@physics.ucla.edu}
\affiliation{Department of Physics and Astronomy, University of California, Los Angeles, CA 90095, USA}
\affiliation{Mani L. Bhaumik Institute for Theoretical Physics,
University of California, Los Angeles, CA 90095, USA}
\affiliation{Center for Frontiers in Nuclear Science, Stony Brook University, Stony Brook, NY 11794, USA}
                   
\author{Xiaohui Liu}
\email{xiliu@bnu.edu.cn}
\affiliation{School of Physics and Astronomy, Beijing Normal University,
and Key Laboratory of Multiscale Spin Physics (Beijing Normal University), Ministry of Education, Beijing 100875, China}
 \affiliation{Southern Center for Nuclear Science Theory (SCNT),
Institute of Modern Physics, Chinese Academy of Science, Huizhou 516000, China}

\author{Sonny Mantry}
\email{sonny.mantry@ung.edu}
\affiliation{Department of Physics and Astronomy, 
                   University of North Georgia,
                   Dahlonega, GA 30597, USA}
\author{Frank Petriello}
\email{f-petriello@northwestern.edu}
\affiliation{Northwestern University, Evanston, IL, USA}

\begin{abstract}
 \noindent
 
We propose a unified framework for a combined global analysis to constrain leading hadronization effects across the 1-Jettiness class of global event shapes for Deep Inelastic Scattering (DIS). We show that for the subclass of jet-based event shapes where the leading jet direction is determined dynamically event-by-event, the leading hadronization effects can acquire a non-trivial dependence on the  hard scattering kinematics. However, this dependence is explicitly calculable, allowing for universality of leading hadronization effects in the 1-Jettiness class. The non-trivial kinematic dependence provides an independent lever arm for simultaneously constraining hadronization effects in the jet-based and  DIS thrust event shapes. This universality, combined with the kinematic lever arm, could allow for including the typically ignored peak region,
where hadronization effects are most severe, in precision extractions of the strong coupling. We demonstrate the need for such a unified treatment of hadronization effects through comparisons of theoretical predictions with simulation data.
\end{abstract}

\maketitle

Global event shapes~\cite{Dasgupta:2003iq,Banfi:2010xy,Stagnitto:2025air,Brandt:1964sa,Farhi:1977sg,Rakow:1981qn,Parisi:1978eg,Catani:1991bd,Stewart:2010tn,Antonelli:1999kx,Dasgupta:2001sh,Dasgupta:2001eq,Dasgupta:2002bw,Dasgupta:2002dc} are infrared and collinear safe observables that quantify the pattern of final state hadronic energy flow through a continuous event shape parameter. They are powerful tools for precision tests of perturbative QCD and the dynamics of hadronization, and precision extractions of the strong coupling~\cite{Becher:2008cf, Abbate:2010xh,Abbate:2012jh,Bell:2023dqs,Benitez:2024nav,Benitez:2025vsp} and the top quark mass~\cite{Fleming:2007qr,Fleming:2007xt,Bachu:2020nqn,Dehnadi:2023msm}.  They group final state particles into beam or jet regions using a set of null reference axes. 
A small event shape value corresponds to   energetic hadrons  restricted to be  collimated along one of the reference axes, corresponding to the reference axes being correlated  with the directions of dominant energy flow emerging from the  hard scattering process.  A large event shape value corresponds to a more spherical pattern of energy flow, uncorrelated with the reference axes. 

The leading nonperturbative hadronization effects occur in the region of  small event shape values, arising from final state soft radiation. They are described by the vacuum matrix element of Wilson lines that encode soft eikonal emissions along the reference axes. For many event shapes, the reference axes are defined with fixed relative angles for all events and across all kinematic bins. For example, the thrust event shape is defined with two reference vectors, always back-to-back along a thrust axis. For such event shapes, the leading nonperturbative hadronization effects are independent of the kinematics of the hard scattering process. In contrast, for jet-based global event shapes such as N-Jettiness~\cite{Stewart:2010tn}, the angles between the  reference axes  vary event-by-event,  correlated with the hard scattering kinematics. Correspondingly, the leading hadronization  effects can acquire a non-trivial kinematic dependence, reflecting clustering effects  that depend on the relative directions of dominant energy flow. The special case of 2-Jettiness~\cite{Becher:2008cf, Abbate:2010xh,Abbate:2012jh,Bell:2023dqs,Benitez:2024nav,Benitez:2025vsp,Fleming:2007qr,Fleming:2007xt,Bachu:2020nqn,Dehnadi:2023msm} in $e^+e^-\to$ hadrons is equivalent to thrust in the dijet limit so that its leading hadronization effects are still independent of the hard kinematics.

In this letter, we propose a unified framework  for a combined global analysis of leading hadronization effects across the  1-Jettiness~\cite{Kang:2012zr,Kang:2013wca,Kang:2013nha,Kang:2013lga,Kang:2014qba,Chu:2022jgs,Cao:2024ota,Ee:2025scz,Chien:2025rbp,Dotson:2026ttc} class of DIS event shapes that include  thrust and jet-based global event shapes.  We show that the leading hadronization effects in jet-based DIS event shapes can acquire a non-trivial dependence on the hard scattering kinematics. This kinematic dependence is explicitly calculable such that the leading hadronization effects in jet-based and  thrust DIS event shapes are described by the same underlying  universal nonperturbative shape function. The non-trivial kinematic dependence provides a new lever arm for better constraining the universal shape function.  This could allow for including the typically ignored peak region,
where hadronization effects are most severe, in precision extractions of the strong coupling.  Previous analyses~\cite{Kang:2012zr,Kang:2013wca,Kang:2013nha,Cao:2024ota,Ee:2025scz}  directly employed different projections of the universal shape function, relevant to the specific  1-Jettiness observable being studied, not conducive for a global analysis and use of the  kinematic lever arm. We derive a more general shape function formula, in terms of the underlying universal shape function,  which reduces to the appropriate projection for a given 1-Jettiness event shape while explicitly encoding any non-trivial kinematic dependence. We also relate the required renormalon subtractions for thrust and jet-based  DIS event shapes up to calculable kinematic  effects. We give predictions at the N$^3$LL level of accuracy, and demonstrate the need for the unified treatment of hadronization effects with comparisons to Pythia~\cite{Sj_strand_2015,Bierlich:2022pfr}. More detailed derivations and  results are given in a companion paper~\cite{companion:2026}.

We study the DIS process $e^-(k)+p(P)\to e^-(k')+J+X$, where $k^\mu,k'^\mu,$ and $P^\mu$ denote the initial electron, final electron, and initial proton four momenta, respectively. Here $J$ is the final state jet, defined within the framework of 1-Jettiness. We work in the center of mass frame, ignoring the electron and proton masses so that $P^\mu=\sqrt{s}n_B^\mu/2$ and $k^\mu=\sqrt{s}\bar{n}_B^\mu/2$,  where $s=(k+P)^2$ is the square of hadronic center of mass energy. The null vectors are $n_B^\mu=(1,\vec{n}_B)$ and $\bar{n}_B^\mu=(1,-\vec{n}_B)$, with $\vec{n}_B=(0,0,1)$ a unit vector  in the direction of the proton beam. The standard DIS kinematic variables are   $Q^2=-q^2=-(k-k')^2$, $x=Q^2/(2P\cdot q)$, and $y=P\cdot q/P\cdot k=Q^2/(xs)$. 
\begin{figure}
\centering
\includegraphics[scale=0.12]{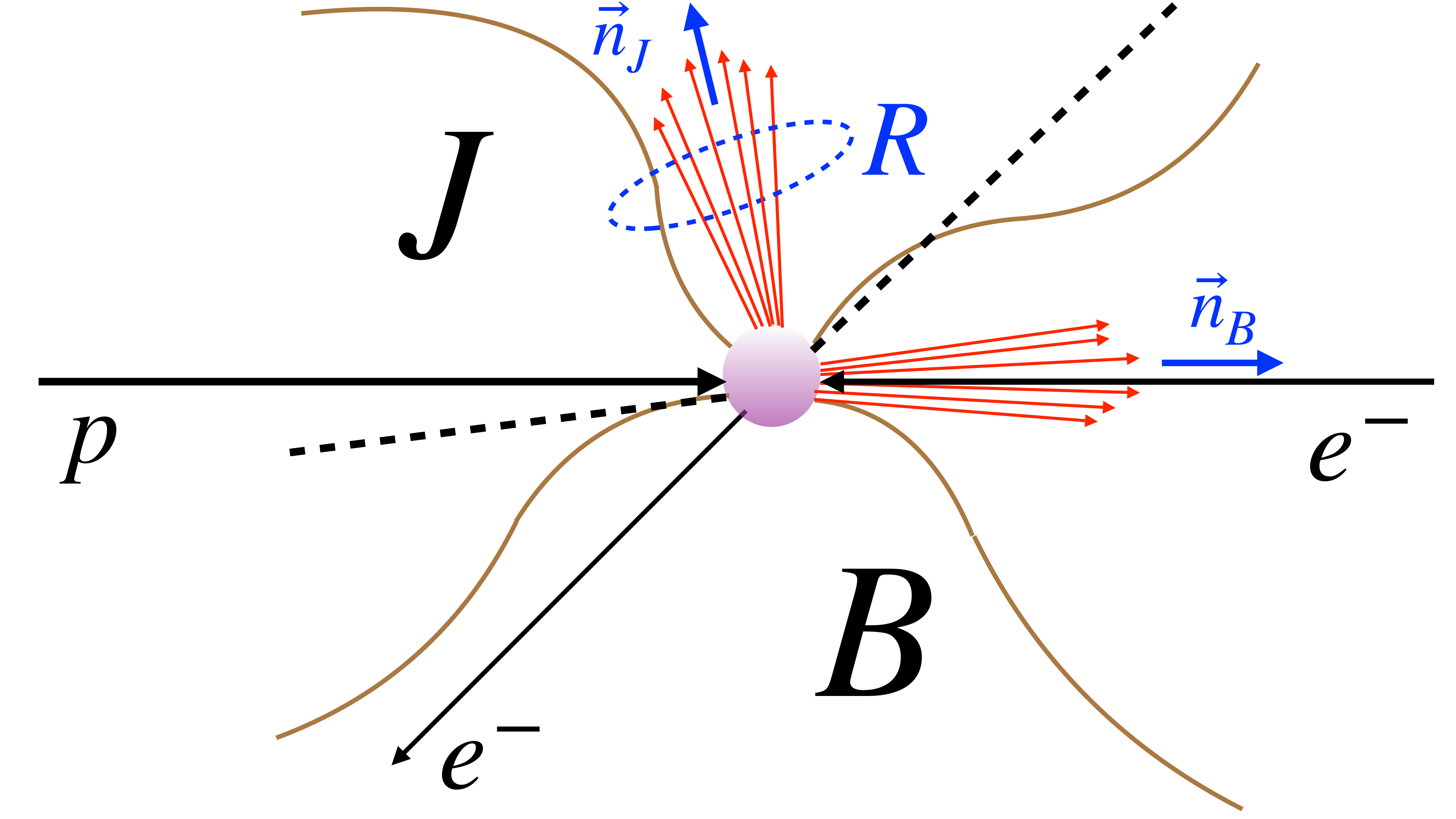}
\caption{1-Jettiness groups  all final state hadrons  into  beam ($B$) or jet ($J$) regions. In the small 1-Jettiness region, shown above, energetic (red) hadrons are restricted to lie along the beam ($\vec{n}_B$) or jet ($\vec{n}_J$) reference axes, with soft (brown) hadrons  unrestricted. The angle $\vec{n}_B\cdot \vec{n}_J$ is determined by the hard kinematics.}
\label{fig:tau1tau1a}
\vspace{-0.6cm}
\end{figure}
We define the class of jet-based 1-Jettiness DIS global event shapes, $\xi$, as 
\bea
\label{tau1andtau1a}
\xi &=& \sum_k \text{min} \Big \{ \frac{2q_B\cdot p_k}{Q_B}, \frac{2q_J\cdot p_k }{Q_J}\Big \}, 
\eea
where the sum is over all final state particles except the final scattered electron.  The  beam  reference vector has the form $q_B^\mu=\omega_B n_B^\mu/2$, and we make the canonical choice $\omega_B=x\sqrt{s}$. The null jet reference vector $q_J^\mu=(K_{J_T} \cosh y_K, \vec{K}_{J_T}, K_{J_T}\sinh y_K)$ is constructed from the  transverse momentum $K_{J_T}$ and rapidity $y_K$ of the leading jet found through a standard jet algorithm, such as the anti-$k_T$ algorithm~\cite{Cacciari:2011ma} with jet radius $R$, which we set to $R=0.5$.  Then we can write $q_J^\mu=\omega_J n_J^\mu/2$ where $\omega_J=2K_{J_T} \cosh y_K$ and $n_J^\mu=(1,\vec{n}_J)$ with $\vec{n}_J$ a unit vector in the direction of the leading jet momentum $\vec{K}_J$. Different choices of $Q_B$ and $Q_J$, on the order of the hard scale, correspond to different definitions of 1-Jettiness. While our conclusions apply to all potential definitions of $\xi$, we focus on the $\tau_1$~\cite{Kang:2012zr,Kang:2013wca} and $\tau_{1a}$~\cite{Kang:2013nha} definitions, expressed in unified notation in terms of $\xi$ as
\bea
\tau_1: \>\>\>\>\xi =\tau_1,  \qquad
\tau_{1a}: \>\>\>\> \xi = Q\tau_{1a}.
\eea
$\xi=\tau_1$ corresponds to the choice $Q_B=\omega_B$ and $Q_J=\omega_J$ and $\xi=Q\tau_{1a}$ corresponds to the choice $Q_B=Q_J=Q=\sqrt{Q^2}$. The 1-Jettiness jet momentum is then defined as $P_J^\mu= \sum_k p_k^\mu \>\theta (\frac{2q_B\cdot p_k}{Q_B} - \frac{2q_J\cdot p_k}{Q_J} )$, corresponding to the sum of the momenta of all particles grouped in the jet region ($J$), as seen in Fig.~\ref{fig:tau1tau1a}. Note that the set of particles within the anti-$k_T$ jet of radius $R$ is only a subset of the particles  in the jet region. We denote   the transverse momentum and rapidity of $P_J^\mu$ as $P_{J_T}$ and $y_J$, respectively, and the typical hard scale in the process as $Q_H\sim \{Q,P_{J_T}\}$.

For small $\xi \ll Q_H$,  energetic radiation is restricted to lie  along the beam or leading jet directions, with unrestricted soft radiation. It effectively acts as a veto on additional jets, generating large Sudakov logarithms of the form $\alpha_s^n\ln ^{2m}(\xi/Q_H)$ for $m\leq n$, requiring resummation. The corresponding factorization formula, derived using the Soft Collinear Effective Theory (SCET)~\cite{Bauer:2000ew,Bauer:2000yr,Bauer:2001ct,Bauer:2001yt,Bauer:2002nz,Beneke:2002ph}, is schematically given by~\cite{Kang:2012zr,Kang:2013wca,Kang:2013nha}
\bea
\label{eq:fac}
d\sigma_{\rm resum}\left[\xi \right ] \sim \sigma_0\> H\otimes B\otimes J \otimes {\cal S},
\eea
where $\sigma_0$ is the Born-level  partonic cross section. We have suppressed  dependence on additional hard kinematic variables such as $\{x,y, Q^2, P_{J_T},y_J\}$ according to which the 1-Jettiness cross section may be binned. The hard ($H$), beam ($B$), jet ($J$), and soft (${\cal S}$) functions describe  the hard scattering, initial state collinear radiation and the initial state PDF, collinear radiation along the jet direction, and final state soft radiation, respectively. Large logarithms in  these functions are minimized at their natural scales $\mu_H\sim Q_H, \mu_B\sim\mu_J\sim \sqrt{Q_H\xi},$ and $\mu_S\sim \xi$, respectively. We use the profile functions~\cite{Berger:2010xi,Stewart:2011cf} in Ref.~\cite{companion:2026} to capture the $\xi$ dependence of these scalings.  Resummation is achieved via the renormalization group to evolve these functions to a common scale at which the cross section is evaluated. The beam function~\cite{Stewart:2009yx} can be matched onto the initial state PDF ($f$) as $B\sim {\cal I}\otimes f$, where ${\cal I}$ describes the perturbative initial state collinear radiation and the PDF is DGLAP evolved to $\mu_B$.
The leading nonperturbative hadronization effects in Eq.~(\ref{eq:fac}) are encoded in the 1-Jettiness soft function
\bea
\label{eq:Sxi}
\hspace{-3mm}{\cal S}(\xi,\mu_S)&=& \int dk_B \hspace{-1mm} \int dk_J \delta(\xi-\lambda_B k_B-\lambda_J k_J) \nn 
\\
&\times& {\cal S}_{\xi} (k_B,k_J,\mu_S), 
\eea
given in terms of the generalized~\cite{Jouttenus:2011wh} hemisphere soft function ${\cal S}_{\xi} (k_B,k_J,\mu_S)$ which divides the final state soft radiation into the jet and beam regions  in Fig.~\ref{fig:tau1tau1a}. The arguments $k_B=\sum_{i\in B} n_B\cdot k_i$ and $k_J=\sum_{i\in J}n_J\cdot k_i$ denote components of the total momentum of soft particles in the beam and jet regions, respectively. For $\xi=\tau_1$ and $\xi=Q\tau_{1a}$, we have $\lambda_{B,J}=1$ and $\lambda_{B,J}=\omega_{B,J}/Q$, respectively.
${\cal S}(\xi,\mu_S)$  depends on the scalar product $n_B \cdot n_J= 1- \tanh y_J$, with $y_J=1/2 \ln \left [Q^2(1-y)/(y^2s)\right ]$.

In contrast, for   DIS thrust~\cite{Antonelli:1999kx,Dasgupta:2001sh,Dasgupta:2001eq,Dasgupta:2002bw,Dasgupta:2002dc,Kang:2013nha,Ee:2025scz}   in the Breit frame, the beam and current jet axes are always back-to-back, along $\hat{z}$ and $-\hat{z}$, respectively. Correspondingly, the soft function that appears in the factorization and resummation formula for DIS thrust is written as~\cite{Kang:2013nha} 
\bea
\label{eq:Sthrust}
{\cal S}^{\rm thrust}(k_S,\mu_S)&=&\int dk_1 \int dk_2 \>\delta (k_S-k_1-k_2)\nn \\
&\times& {\cal S}_{\rm hemi.}(k_1,k_2,\mu_S), 
\eea
in terms of the hemisphere soft function ${\cal S}_{\rm hemi.}(k_1,k_2,\mu_S)$ which groups final state soft particles into back-to-back hemispheres. The arguments  $k_1=\sum_{i\in 1} n_1\cdot k_i$ and $k_2=\sum_{i\in 2}n_2\cdot k_i$ denote the total light-cone momentum components of  soft particles in hemispheres 1 and 2, respectively.  The null reference vectors are $n_1^\mu=(1,\hat{z})$ and $n_2^\mu=(1,-\hat{z})$, with constant scalar product  $n_1\cdot n_2=2$.
 
 Using boost invariance of soft Wilson lines  and the boost transformation properties of  $k_B$ and $k_J$,  Ref.~\cite{Kang:2013nha} proved the relation 
 \bea
\label{eq:Sgenhemi}
{\cal S}_{\xi}(k_B,k_J,\mu_S) =\frac{1}{R_BR_J} {\cal S}_{\rm hemi.}\left(\frac{k_B}{R_B},\frac{k_J}{R_J},\mu_S\right),
\eea
where the $R_B$ and $R_J$ rescaling factors are used to define a new set of transformed reference null vectors $n_B'= n_B/R_B$ and  $n_J'= n_J/R_J$
such that they satisfy the same condition as for  DIS thrust, $n_B'\cdot n_J'=n_1\cdot n_2=2$. For $\tau_1$~\cite{Cao:2024ota}, we have $R_B^2 = R_J^2 = n_B\cdot n_J/2$ and for  $\tau_{1a}$~\cite{Kang:2013nha},  $R_{B}^2 = \omega_{J}n_{B}\cdot n_{J}/(2\omega_{B})$ and $R_{J}^2 = \omega_{B}n_{B}\cdot n_{J}/(2\omega_{J})$. Using Eq.~(\ref{eq:Sgenhemi}) in Eq.~(\ref{eq:Sxi}), the 1-Jettiness soft function  can be expressed in terms of ${\cal S}_{\rm hemi.}$   as
\bea
\label{eq:Stauxi}
{\cal S}(\xi,\mu_S) &=& \int dk_B \int dk_J \>\delta(\xi - {\cal R}_B k_B -{\cal R}_J k_J). \nn \\
&\times&{\cal S}_{\rm hemi.}(k_B,k_J,\mu_S) ,
\eea
where ${\cal R}_{B,J}=\lambda_{B,J} R_{B,J}$.
Eqs.~(\ref{eq:Stauxi}) and (\ref{eq:Sthrust}) show that the soft functions for $\tau_1$, $\tau_{1a}$, and DIS thrust correspond to different projections of the same underlying hemisphere soft function  ${\cal S}_{\rm hemi.}(k_1,k_2,\mu_S)$. Any non-trivial kinematic dependence  in ${\cal S}(\xi,\mu_S)$ is explicitly encoded in the  ${\cal R}_{B,J}$ factors. 
This is a key result that implies that hadronization effects, including hadron mass effects~\cite{Mateu:2012nk}, be incorporated  in a unified manner through ${\cal S}_{\rm hemi.}(k_1,k_2,\mu_S)$ for $\tau_1$, $\tau_{1a}$, DIS thrust, or other definitions of $\xi$. Previous analyses~\cite{Kang:2012zr,Kang:2013wca,Kang:2013nha,Cao:2024ota,Ee:2025scz}  have treated hadronization effects for each 1-Jettiness observable separately by directly working with the projections  ${\cal S}(\tau_1,\mu_S)$, ${\cal S}(Q\tau_{1a},\mu_S)$, or ${\cal S}^{\rm thrust}(k_s,\mu_S)$, preventing a clear interpretation in the context of a combined global analysis.

For $\tau_{1a}$, one can explicitly check that  ${\cal R}_{B,J}=1$ in the $\xi\ll Q_H$ limit. This implies that $\tau_{1a}$ corresponds to a special case where the kinematic dependence exactly  cancels  so that it has the same functional dependence on ${\cal S}_{\rm hemi.}$ as ${\cal S}^{\rm thrust}$, as first shown in Ref.~\cite{Kang:2013nha}. However, for $\tau_1$,  ${\cal R}_{B,J}=R_{B,J}$, leading to a non-trivial kinematic dependence. 
 These results are also consistent with  fixed-order perturbative calculations~\cite{Jouttenus:2011wh,Boughezal:2015eha} of ${\cal S}(\xi,\mu_S)$. A key feature of the unified treatment of hadronization effects through ${\cal S}_{\rm hemi.}$, is that it allows the use of $\tau_1$ as an independent kinematic lever arm for simultaneously constraining hadronization effects in $\tau_1,\tau_{1a},$ and DIS thrust.

Hadronization effects in ${\cal S}_{\rm hemi.}(k_1,k_2,\mu_S)$ are incorporated via a convolution with a  shape function~\cite{Hoang:2007vb} as
\bea
\label{eq:hemiconvgap}
 &&{\cal S}_{\rm hemi.}(k_1,k_2,\bar{\Delta},\mu_S) = \int dk_1' \int dk_2'  \\
 &\times&  {\cal S}_{\rm hemi.}^{\rm part.}(k_1-k_1',k_2-k_2',\mu_S) \>{\cal S}_{\rm hemi.}^{\rm mod.}(k_1'-\bar{\Delta},k_2'-\bar{\Delta}) . \nn
\eea
where the dependence on the gap parameter $\bar{\Delta}$ is made explicit.    ${\cal S}_{\rm hemi.}^{\rm part.}(k_1,k_2,\mu_S)$ denotes the partonic hemisphere soft function, calculated in perturbation theory and contains the renormalization scale dependence.  The model shape function satisfies the normalization $\int dk_1\int dk_2 \>{\cal S}_{\rm hemi.}^{\rm mod.}(k_1,k_2) =1$, and is chosen to have support only for $k_{1,2}\geq 0$, with an exponential fall off for $k_{1,2}\gg \Lambda_{\rm QCD}$. The convolution structure then implies  that  ${\cal S}_{\rm hemi.}(k_1,k_2,\bar{\Delta}=0,\mu_S)$ has non-zero support for $k_{1,2}\geq 0$, corresponding to the partonic threshold.  A non-zero gap parameter $\bar{\Delta}\sim \Lambda_{\rm QCD}$  encodes the QCD mass gap, corresponding to the hadronic threshold with non-zero support for $k_{1,2}\geq \bar{\Delta}$ in ${\cal S}_{\rm hemi.}(k_1,k_2,\bar{\Delta},\mu_S)$.

\begin{figure}
\centering
\includegraphics[scale=0.55]{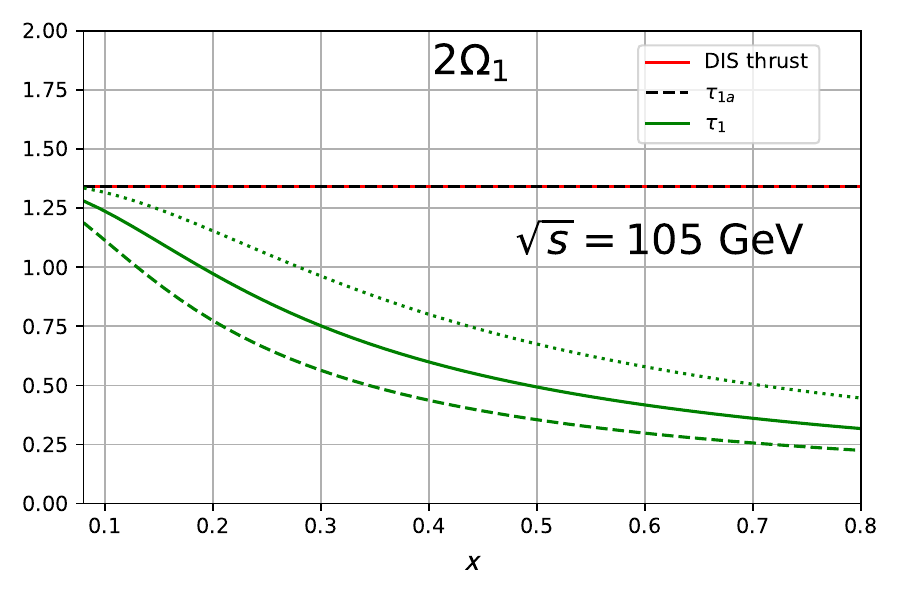}
\vspace{-0.5cm}
\caption{Kinematic dependence of $2\Omega_1$  for DIS thrust (red),  $\tau_1$ (green), and $\tau_{1a}$ (black).  There is no $(x,Q^2)$ kinematic dependence for DIS thrust and $\tau_{1a}$. For $\tau_1$,  the dashed, solid, and dotted green curves correspond to $Q^2=[200,400,800]$ GeV$^2$, respectively. }
\label{fig:Omega1}
\vspace{-0.5cm}
\end{figure} 
It is well-known~\cite{Gardi:2000yh,Hoang:2007vb,Hoang:2008fs} that ${\cal S}_{\rm hemi.}^{\rm part.}$  in the $\overline{\rm MS}$ scheme has an infrared renormalon, leading  to an ${\cal O}(\Lambda_{\rm QCD})$ ambiguity in the  partonic threshold. Through Eq.~(\ref{eq:hemiconvgap}), this ambiguity translates into an ${\cal O}(\Lambda_{\rm QCD})$ ambiguity in the gap parameter $\bar{\Delta}$ itself. Thus, the renormalon in ${\cal S}_{\rm hemi.}^{\rm part.}$ can be subtracted by  writing~\cite{Hoang:2008fs,Abbate:2010xh,Jain:2008gb,Hoang:2008yj} the gap parameter as $\bar{\Delta} =\Delta(R,\mu_S) + \delta(R,\mu_S) $,
where $\delta(R,\mu_S)$ is a perturbative series in the R-gap scheme~\cite{Hoang:2008fs}, proportional to $R$, and
containing the same renormalon as in ${\cal S}_{\rm hemi.}^{\rm part.}$. The scale $R$ corresponds to the renormalon subtraction scheme choice, and is naturally of size $R\sim \Lambda_{\rm QCD}$ corresponding to a ${\cal O}(\Lambda_{\rm QCD})$ shift in $\bar{\Delta}$. This defines $\Delta(R,\mu_S)$ as a renormalon-free gap parameter, obeying evolution equations in $R$-$\mu_S$ space determined by $\delta(R,\mu_S)$, since $\bar{\Delta}$ is constant. Large logarithms of  $\mu_S/R$ are resummed by evolving in $R$-$\mu_S$ space to a point where $R\sim \mu_S$. The terms up to ${\cal O}(\alpha_s^2)$ in $\delta(R,\mu_S)$, needed for N$^3$LL resummation, are now known~\cite{Hoang:2008fs,Kang:2015moa,Ee:2025scz}.
\begin{figure}[htbp]
    \centering
    \includegraphics[width=0.43\textwidth]{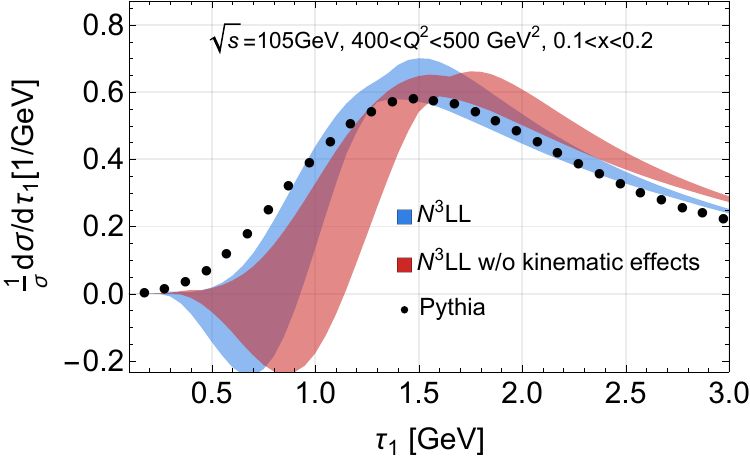}

    \includegraphics[width=0.45\textwidth]{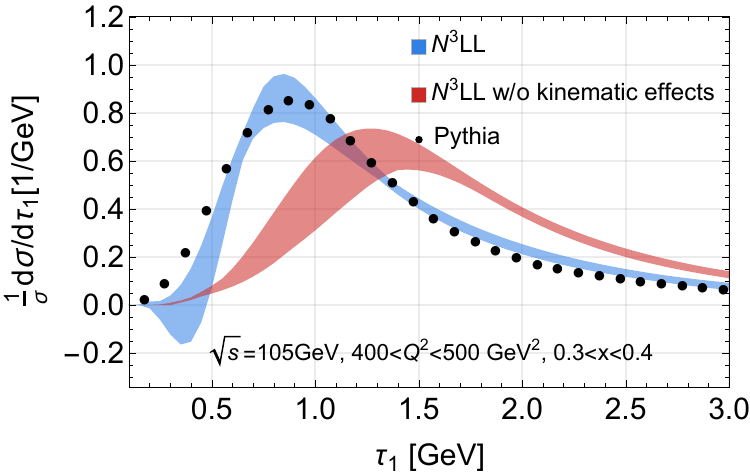}

    \includegraphics[width=0.45\textwidth]{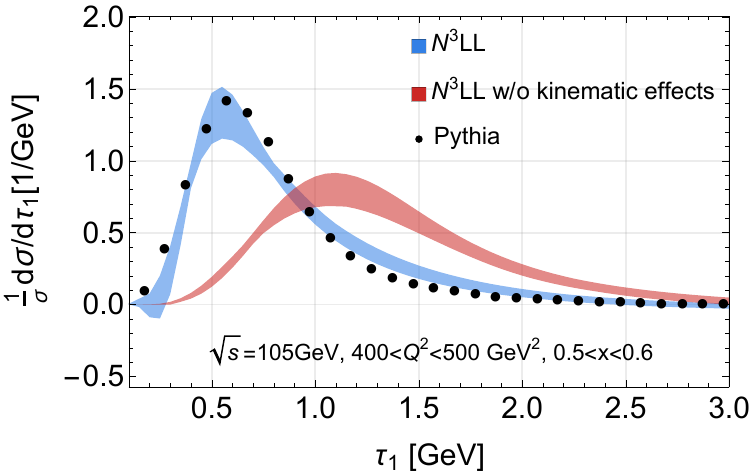}
    \caption{Normalized hadron-level N$^3$LL  $\tau_1$-distributions with (blue) and without (pink) kinematic effects in hadronization, compared to Pythia data (black dots).}
    \label{fig:tau1}
\end{figure}
\begin{figure}[htbp]
   \vspace{0.5cm}
    \centering
    \includegraphics[width=0.43\textwidth]{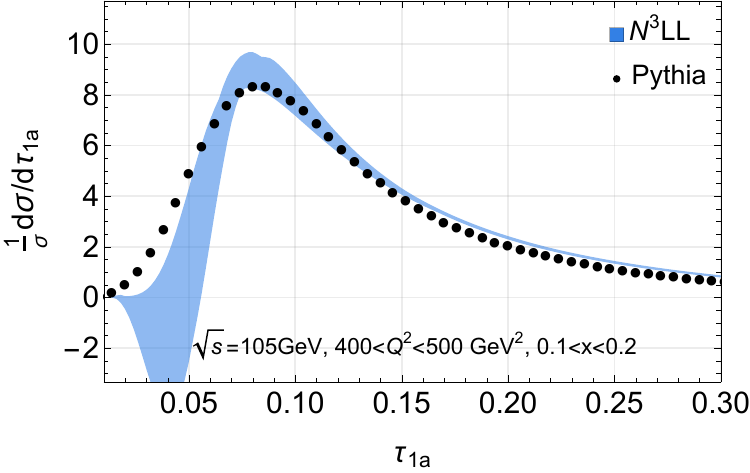}

    \includegraphics[width=0.45\textwidth]{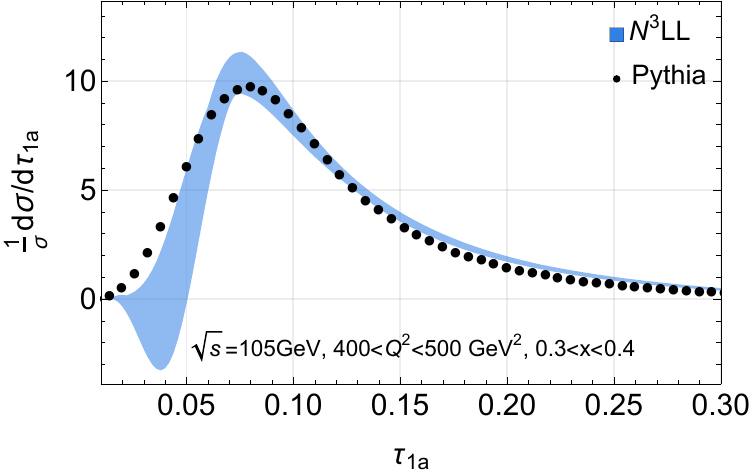}

    \includegraphics[width=0.45\textwidth]{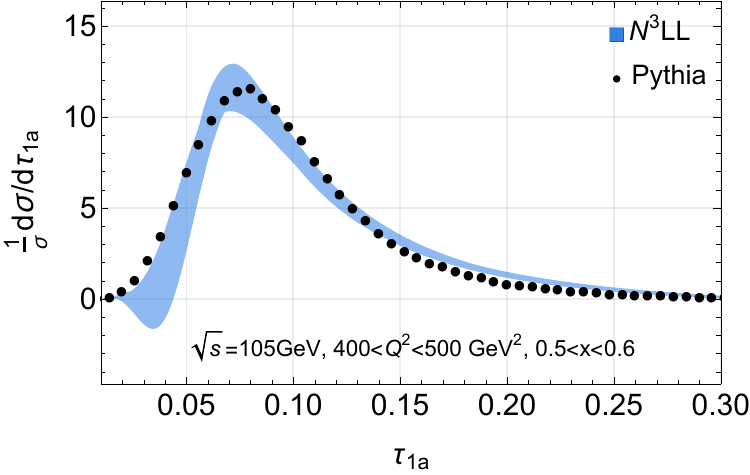}
    \caption{Normalized hadron-level N$^3$LL  $\tau_{1a}$-distributions  (blue)  compared to Pythia data (black dots).}
    \label{fig:tau1a}
    \vspace{-0.3cm}
\end{figure}
Using Eq.~(\ref{eq:hemiconvgap}) in Eq.~(\ref{eq:Stauxi}), the renormalon-subtracted 1-Jettiness soft function takes the form
\bea
\label{eq:SxiRS}
{\cal S}(\xi,\delta,\mu_S) &=& \int du\>{\cal S}^{\rm part.}\left (\xi-u - \left({\cal R}_B+{\cal R}_J\right)\delta,\mu_S\right) \nn \\
&\times& F^{\rm mod.}(u,{\cal R}_B,{\cal R}_J,\Delta) .
\eea
${\cal S}^{\rm part.}\left (\xi,\mu_S\right)$ is the partonic 1-Jettiness soft function without renormalon subtractions, obtained by the replacement ${\cal S}_{\rm hemi.}\to {\cal S}_{\rm hemi.}^{\rm part.}$ in Eq.~(\ref{eq:Stauxi}). Thus, there is an implicit ${\cal R}_{B,J}$-dependence in ${\cal S}^{\rm part.}$ in addition to  the renormalon subtraction term $\left({\cal R}_B+{\cal R}_J\right)\delta$  in its argument in Eq.~(\ref{eq:SxiRS}). Note that $\delta=\delta(R,\mu_S)$ is the same renormalon subtraction series as for the DIS thrust hemisphere soft function.  The shape function $F^{\rm m od.}(u,{\cal R}_B,{\cal R}_J,\Delta)$  is given by
\bea
\label{eq:FmodRS}
\hspace{-4.5mm}F^{\rm mod.} \hspace{-1mm}= \hspace{-1mm}\int \frac{d\zeta}{2{\cal R}_B {\cal R}_J} {\cal S}_{\rm hemi.}^{\rm mod.}\left(\frac{u+\zeta}{2{\cal R}_B}-\Delta,\frac{u-\zeta}{2{\cal R}_J}-\Delta\right),
\eea
with the limits of integration  determined by  ${\cal S}_{\rm hemi.}^{\rm mod.}(k_1,k_2)$ being non-zero only for $k_{1,2}\geq 0$.
Eqs.~(\ref{eq:SxiRS}) and (\ref{eq:FmodRS}) are  key results,  relating  hadronization effects and renormalon subtractions in $\tau_1,\tau_{1a}$, and DIS thrust. The non-trivial kinematic dependence of hadronization effects is contained in the ${\cal R}_{B,J}$ factors in $F^{\rm mod.}$.  Since ${\cal R}_{B,J}=1$ for $\tau_{1a}$, its shape function is identical to that of DIS thrust, $F^{\rm m od.}(u,{\cal R}_B=1,{\cal R}_J=1,\Delta)$. Using Eq.~(\ref{eq:SxiRS})  in Eq.~(\ref{eq:fac}), the cross section takes the form
\bea
\label{eq:fac_intermediate_1}
d\sigma_{\rm resum}  \left [ \xi, \delta \right]
&=& \hspace{-1.5mm} \int du \>d\sigma_{\rm resum}^{\rm part.} \left [\xi- u - ({\cal R}_B +{\cal R}_J) \delta\right ] \nn \\
&\times& F^{\rm mod.}(u, {\cal R}_B,{\cal R}_J,\Delta) ,
\eea
where $d\sigma_{\rm resum}^{\rm part.}\left [\xi \right ]$ is the partonic cross section, corresponding to Eq.~(\ref{eq:fac}) computed with the partonic soft function ${\cal S}^{\rm part.}$.
Since $F^{\rm mod.}$ peaks near $u\sim \Lambda_{\rm QCD}$, the tail region,  $\Lambda_{QCD} \ll \xi \ll Q_H$, allows an operator product expansion (OPE), leading to the well-known~\cite{Korchemsky:1994is,Dokshitzer:1995zt,Dokshitzer:1995qm} result that leading nonperturbative effects in the tail region are given by a simple shift in the partonic cross section:  $d\sigma_{\rm resum}\left [\xi, \delta\right ]  \to d\sigma_{\rm resum}^{\rm part.}\left [\xi -({\cal R}_B+{\cal R}_J)\delta- 2 \Omega_1\right ]$,
where the renormalon-subtracted first moment $2\Omega_1=2\Omega_1(R,\mu_S,{\cal R}_B,{\cal R}_J) $ is given by
\bea
\label{eq:moment}
2\Omega_1 = \int du \>u\> F^{\rm mod.}(u, {\cal R}_B,{\cal R}_J,\Delta) .
\eea
The new feature here is that  the $2\Omega_1$  shift could vary across kinematic bins. 
For $F^{\rm mod.}$ in Eq.~(\ref{eq:FmodRS}), we choose the model parameterization~\cite{Korchemsky:2000kp} for $S^{\rm mod.}_{\rm hemi.}(k_1,k_2)$ as
\bea
\label{eq:shapehemi}
\hspace{-0.5cm}S^{\rm mod.}_{\rm hemi.} = \frac{{\cal N}}{\Lambda^2}\left ( \frac{k_1 k_2}{\Lambda^2}\right )^{a-1} \hspace{-0.5cm}{\rm exp}\left [\frac{-k_1^2 -k_2^2 - 2b k_1 k_2}{\Lambda^2} \right ], 
\eea
where ${\cal N}$ is a normalization factor, and we set $a=2.0,  b=0.0, \Lambda= 0.4 \>{\rm GeV}$. Following  Ref.~\cite{Ee:2025scz}, we set  $\Delta(R_\Delta,\mu_\Delta)=0.05\> {\rm GeV}$ at the input scales $ R_\Delta=\mu_\Delta=2 \>{\rm GeV}$, and evolve to the point $R=0.85 \mu_S$. Fig.~\ref{fig:Omega1} shows the non-trivial kinematic dependence of $2\Omega_1$ for $\tau_1$, compared to its constant identical value for both DIS thrust and $\tau_{1a}$. In Fig.~\ref{fig:tau1}, we show theoretical predictions (blue bands) for $\tau_1$, compared to Pythia (black dots), at the N$^3$LL level of accuracy for a fixed $Q^2$-bin in three different $x$-bins.  We also show the effect of turning off (pink bands) the kinematic dependence of hadronization effects, corresponding to setting ${\cal R}_B={\cal R}_J=1$ in $F^{\rm mod}$. We see a clear shift of the peak  to the left  for increasing $x$-bins, as expected from the decrease of  $2\Omega_1$ with $x$ in Fig.~\ref{fig:Omega1}. Fig.~\ref{fig:tau1a} shows the corresponding results for $\tau_{1a}$, 
exhibiting a more stable peak position across $x$-bins, as expected from Fig.~\ref{fig:Omega1}.

\textit{Conclusions:} We have shown that leading hadronization effects for the $\tau_1, \tau_{1a}$ and thrust DIS event shapes  can be treated in a unified manner. A key feature of this approach is that the non-trivial but calculable kinematic dependence of hadronization effects in $\tau_1$ can be used as an independent lever arm to simultaneously constrain hadronization effects in these observables. Such a unified treatment is essential for a consistent description of Pythia data across different observables and kinematic bins. This approach can also facilitate precision extractions of the strong coupling by further constraining leading hadronization effects.

{\textit{Acknowledgements:} 
H. C. is partially supported by a CFNS Joint Postdoctoral Fellowship. Z.~K. is supported by the National Science Foundation under grant No.~PHY-2515057. H. C. and F. P. are supported by the U.S. Department of Energy, Office of High Energy Physics, under contract No.~DE-SC0010143. X. L. is supported by the National Natural Science Foundation of China under Grant No.~12547109 and Fundamental Research Funds for the
Central Universities, Beijing Normal University.
This research was supported in part through the computational resources and staff contributions provided for the Quest high performance computing facility at Northwestern University which is jointly supported by the Office of the Provost, the Office for Research, and Northwestern University Information Technology. Z.~K. and  X.~L. would like to thank the Erwin-Schr\"odinger International Institute for Mathematics and Physics at the University of Vienna for partial support during the Programme `New Paradigms for Harnessing Quantum Field Theory at Colliders', July 27 – August 28, 2026. S.M. thanks Jefferson Lab for their hospitality and support through the Visiting Faculty Program (VFP) during which part of this work was carried out.

\bibliographystyle{apsrev4-1}
\bibliography{main-short}
\end{document}